\documentclass[runningheads]{llncs}
\usepackage{graphicx}
\usepackage{multirow}
\usepackage{tikz}
\usepackage{comment}
\usepackage{amsmath,amssymb} 
\usepackage{color}
\usepackage{hyperref}

\usepackage[accsupp]{axessibility}  

\begin{document}
\pagestyle{headings}
\mainmatter
\def\ECCVSubNumber{22}  

\title{
A Data-Efficient Deep Learning Framework for Segmentation and Classification of Histopathology Images
} 

\titlerunning{ECCV-22 submission ID \ECCVSubNumber} 
\authorrunning{ECCV-22 submission ID \ECCVSubNumber} 
\author{Anonymous ECCV submission}
\institute{Paper ID \ECCVSubNumber}

\titlerunning{Data-Efficient Deep Learning Framework}
%
\author{Pranav Singh\inst{1} \and
Jacopo Cirrone\inst{2}}
\authorrunning{P.Singh \& J.Cirrone}
%
\institute{Department of Computer Science\\
  Tandon School of Engineering\\
  New York University
  New York, NY 11202, USA \\
  \email{ps4364@nyu.edu}
  \and
Center for Data Science\\
  New York University\\
  and Colton Center for Autoimmunity\\
  NYU Grossman School of Medicine\\
  New York, NY 10011 \\
\email{cirrone@courant.nyu.edu}\\}

\maketitle

\begin{abstract}
The current study of cell architecture of inflammation in histopathology images commonly performed for diagnosis and research purposes excludes a lot of information available on the biopsy slide. In autoimmune diseases, major outstanding research questions remain regarding which cell types participate in inflammation at the tissue level, and how they interact with each other. While these questions can be partially answered using traditional methods, artificial intelligence approaches for segmentation and classification  provide a much more efficient method to understand the architecture of inflammation in autoimmune disease, holding great promise for novel insights. In this paper, we empirically develop deep learning approaches that use dermatomyositis biopsies of human tissue to detect and identify inflammatory cells. Our approach improves classification performance by 26\% and segmentation performance by 5\%. We also propose a novel post-processing autoencoder architecture that improves segmentation performance by an additional 3\%.
\keywords{Deep Learning, Computer Vision, Medical Image Analysis, Autoimmune diseases, Histopathology Images. }
\end{abstract}

\section{Introduction}
Our understanding of diseases and their classification has improved multifold with developments in medical science. However, our understanding of autoimmune diseases continues to be incomplete, missing vital information. No mechanism is in place to systematically collect data about the prevalence and incidence of autoimmune diseases (as it exists for infectious diseases and cancers). This deficiency is because we lack a comprehensive and universally acceptable list of autoimmune diseases. The most cited study in the epidemiology of autoimmune diseases \cite{jacobson1997epidemiology}, estimates that autoimmune diseases, combined, affects about 3\% of the US population or 9.9 million US citizens. A number comparable to the 13.6 million (US citizens) affected by cancer, which affects almost 4\% of the population. Nevertheless, cancer has been widely studied in medical science and at the intersection of medical science and artificial intelligence. The study of autoimmune diseases is not only critical because they affect a sizeable portion of the population, but because we don't have a complete understating of their etiology and treatment. Other compelling reasons to study these diseases is their increasing prevalence \cite{dinse2020increasing,ijcd2015348} and their further increase with the recent COVID-19 pandemic \cite{galeotti2020autoimmune,ehrenfeld2020covid}.\\

Major outstanding research questions exist for autoimmune diseases regarding the presence of different cell types and their role in inflammation at the tissue level. In addition to studying preexisting patterns for different cell interactions, the identification of new cell occurrence and interaction patterns will help us better understand the diseases. While these patterns and interactions can be partially answered using traditional methods, artificial intelligence approaches for segmentation and classification tasks provide a much more efficient and quicker way to understand these architectures of inflammation in autoimmune disease and hold great promise for novel insights. The application of artificial intelligence for medical image analysis has also seen a rapid increase, propelled by the increase in performance and efficiency of such architectures. However, even with these developments mentioned previously, the application of artificial intelligence in autoimmune biopsy analysis has not received the same attention as others. Firstly, autoimmune diseases are highly underrepresented because of significantly fewer data available for aforementioned reasons. Secondly, even within the few existing studies on the application of artificial intelligence for autoimmune disease analysis, dermatomyositis has received significantly less attention.
Most research has focused on psoriasis, rheumatoid arthritis, lupus, scleroderma, vitiligo, inflammatory bowel diseases, thyroid eye sisease, multiple sclerosis sisease, and alopecia \cite{tsakalidou2022computer}. We also observe that most of these approaches are based on older techniques and architectures that do not have open-source code to allow more researchers to expand their investigations into this area.\\

To help bridge this gap, we aim to draw more attention in this paper to autoimmune diseases, specifically to dermatomyositis. With this paper: (i) we improve upon the existing method for classification and segmentation of autoimmune disease images \cite{VANBUREN2022113233} with 26\% improvement for classification and 5\%  for segmentation, (ii) we propose an \textbf{A}utoencoder for \textbf{P}ost-\textbf{P}rocessing (APP) and using image reconstruction loss improve segmentation performance by a further 3\% as compared to (i), and (iii) based on these experimentation, we make recommendations for future researchers/practitioners to improve the performance of architectures and understanding of autoimmune diseases.
All the aforementioned contributions have been implemented in PyTorch and are publicly available at \url{https://github.com/pranavsinghps1/DEDL}.

\section{Background}

\subsection{Application of Artificial intelligence for autoimmune diseases}

Researchers in \cite{tsakalidou2022computer} conducted an in-depth study of the application of computer vision and deep learning in autoimmune disease diagnosis. Based on their study, we found a common trend among the datasets within autoimmune medical imaging - most of the datasets used are extremely small, with a median size of 126 samples. This also correlated with the findings of \cite{stafford2020systematic}, wherein they also mentioned the median dataset size available for autoimmune diseases ranged between 99-540 samples. Medical imaging datasets tend to be smaller than natural image datasets, and even within medical datasets, autoimmune datasets are comparably smaller than the datasets of diseases with similar prevalence. For example, the prevalence of cancer is around 4\% compared to the prevalence of autoimmune diseases at around 3\%. However, cancer datasets of sizes ranging from a few thousand samples are readily available as opposed to that of autoimmune, where the median dataset size is between 99-540 samples. Another difference is that most of the autoimmune disease datasets are institutionally restricted.
\\
In addition, there are few studies on the application of artificial intelligence in autoimmune diseases. In 2020, \cite{stafford2020systematic} conducted a systematic review of the literature and relevant papers at the intersection of artificial intelligence and autoimmune diseases with the following exclusion criteria: studies not written in English, no actual human patient data included, publication prior to 2001, studies that were not peer-reviewed, non-autoimmune disease comorbidity research and review papers. Only 169 studies met the criteria for inclusion. On further analyzing these 169 studies, only a small proportion of studies 7.7\% (13/169) combined different data types. Cross-validation, combined with independent testing set for a more robust model evaluation, occurred only in 8.3\% (14/169) of papers.
\\
In 2022, \cite{tsakalidou2022computer} studied the usage of computer vision in autoimmune diseases, its limitations and the opportunities that technology offers for future research. Out of the more than 100 classified autoimmune diseases, research work has mostly focused ten diseases (psoriasis, rheumatoid arthritis, lupus, scleroderma, vitiligo, inflammatory bowel diseases, thyroid eye disease, multiple sclerosis disease and alopecia). \cite{VANBUREN2022113233} is the first, to the best of our knowledge, to apply and study artificial intelligence for medical image analysis of dermatomyositis - an autoimmune disease that has not been studied in much detail. We used the same dataset as used by \cite{VANBUREN2022113233} and propose innovative techniques and architectures to improve performance.
\subsection{Segmentation}
Medical image segmentation, i.e., automated delineation of anatomical structures and other regions of interest (ROIs) paradigms, is an important step in computer-aided diagnosis; for example it is used to extract key quantitative measurements and localize a diseased area from the rest of the slide image. Good segmentation requires the object to see fine picture and intricate details at the same time. The same encoder-decoder architectures have been favored that often use different techniques (e.g., feature pyramid network, dilated networks and atrous networks) to help increase the receptive field of the architecture.
When it comes to medical image segmentation, UNet \cite{ronneberger2015u} has been the most cited and widely used network architecture. It uses an encoder and decoder architecture with skip connections to learn the segmentation masks. An updated version of UNet was introduced by \cite{zhou2018unet++} called UNet++ that is essentially a nested version of UNet. The encoder is a feature extractor that down-samples the input, the decoder then consecutively up-samples to learn the segmentation mask for an input image.
We added channel level attention with the help of squeeze and excitation blocks as proposed in \cite{hu2018squeeze}. A squeeze and excitation is basically a building block that can be easily incorporated with CNN architecture. Comprised of a bypass that emerges after normal convolution, this is where the squeeze operation is performed. Squeezing basically means compressing each two-dimensional feature map until it becomes a real number. This is followed by an excitation operation that generates a weight for each feature channel to model relevance. Applying these weights to each original feature channel, the importance of different channels can be learned.

\subsection{Classification}
Classification is another important task for medical image analysis. CNNs have been the de facto standard for classification task. The adoption of Transformers for vision tasks from language models have been immensely beneficial. This gain in performance could be the result of Transformer’s global receptive field as opposed to limited receptive field on CNNs. Although CNNs have inductive priors that make them more suited for vision tasks, Transformers learn them over the training period. Recently, there has been increased interest in combining the abilities of Transformers and CNNs. Certain CNNs have been trained the way Transformers have been trained as in \cite{liu2022convnet}, similarly the introduction of CNN type convolution has been incorporated in Transformers\cite{liu2021swin}.Within medical image analysis, Transformers with their global receptive field could be extremely beneficial as this can help us learn features that CNNs with their limited receptive field could have missed. We start with the gold-standard of CNNs - ResNet-50 and the ResNet family \cite{he2015deep} of architectures. 
A ResNet-18 model can be scaled up to make ResNet-200. In most cases, this yields a better performance. But this scaling is very random as some models are scaled depth-wise and some are scaled width-wise. This problem was addressed with \cite{tan2019efficientnet}, in which the concept of compound scaling was used.
They proposed a family of models with balanced scaling that also improved overall model performance. In ResNet-like architectures, batch normalization is often applied at the residual branch. This stabilizes the gradient and enables training of significantly deeper networks. Yet computing batch-level statistics is an expensive operation. To address these issues a set of NFnets (Normalizer-Free Networks) were published \cite{brock2021high}. Instead of using batch normalization, NFnets use other techniques to create batch-normalization like effect such as modified residual branches and convolutions with scaled weight standardization and adaptive gradient clipping. 
Transformers have a global receptive field, as opposed to CNN. \cite{dosovitskiy2020image} introduced Vision Transformers (ViT) adapted from NLP, and since then they have improved upon many benchmarks for vision tasks. Hence, we study the effect of using Transformers as classifiers on the autoimmune dataset. 
Within Transformers, we start by looking at vision transformers introduced in \cite{dosovitskiy2020image}. Unlike CNNs, Transformer first split each image into patches by a patch module. These patches are then used as ”tokens”. We then examine the Swin transformer family \cite{liu2021swin}. Swin transformers use a hierarchical approach with shifted windows. The shifted windows scheme brings greater efficiency by limiting self-attention computation to non-overlapping windows and allows cross-window connection. It also first splits an input RGB image into non-overlapping image patches by a patch splitting module, like ViT\cite{dosovitskiy2020image}.

\section{Methodology}
\subsection{Segmentation}

For segmentation, our contribution is twofold. Firstly, we improved upon existing approaches by reducing blank image tiling, adding channel level attention in decoder, by using squeeze and excitation blocks to better map the importance of different channels and by using a pixel normalized cross-entropy loss. This improved performance by 5\% over existing approach on the same dataset by \cite{VANBUREN2022113233}, secondly, we introduce a novel post-processing Autoencoder which further improved the performance of the segmentation architecture as compared to the first step by 3\%.

We used the same dermatomyositis dataset as used in \cite{VANBUREN2022113233}. We started our experimentation by using the same metrics, image tiling and splits as their work, to make comparison easier. Once we surpassed their benchmark, we changed a few things as described in Section 5.1.


For segmentation benchmark as mentioned already, we based our study around one of the most widely-cited networks for biomedical image segmentation - UNet and a nested version of UNet - UNet++. These architectures are widely used not only in general segmentation tasks but also in biomedical segmentation. UNet has also been used in previous autoimmune segmentation tasks \cite{VANBUREN2022113233,DASH2019226}.


\subsubsection{Intuition for the Autoencoder Post-Processing architecture.}

Traditionally for training segmentation architectures, the output from the decoder is compared against ground truth with a loss function, in our case, we used a cross-entropy loss function. Increasing the model size should help the model with more extracted features from the input but based on our experimentation, this is not the case Table~\ref{table:aechnage}. So to provide the segmentation architecture with meaningful insight to improve learning, we provide additional feedback on ”How easy is it to reconstruct the ground truth mask with the predicted mask?”. To do so, we used simple encoder-decoder architecture with cross entropy loss. This post-processing autoencoder takes the output of the segmentation architecture as input and then computes this reconstruction loss, which the leading segmentation architecture then uses to improve learning. We expanded more on the experiments and study the results of adding these autoencoders in Section 5.1.


\begin{figure}[!htb]
\centering
\includegraphics[height=8cm]{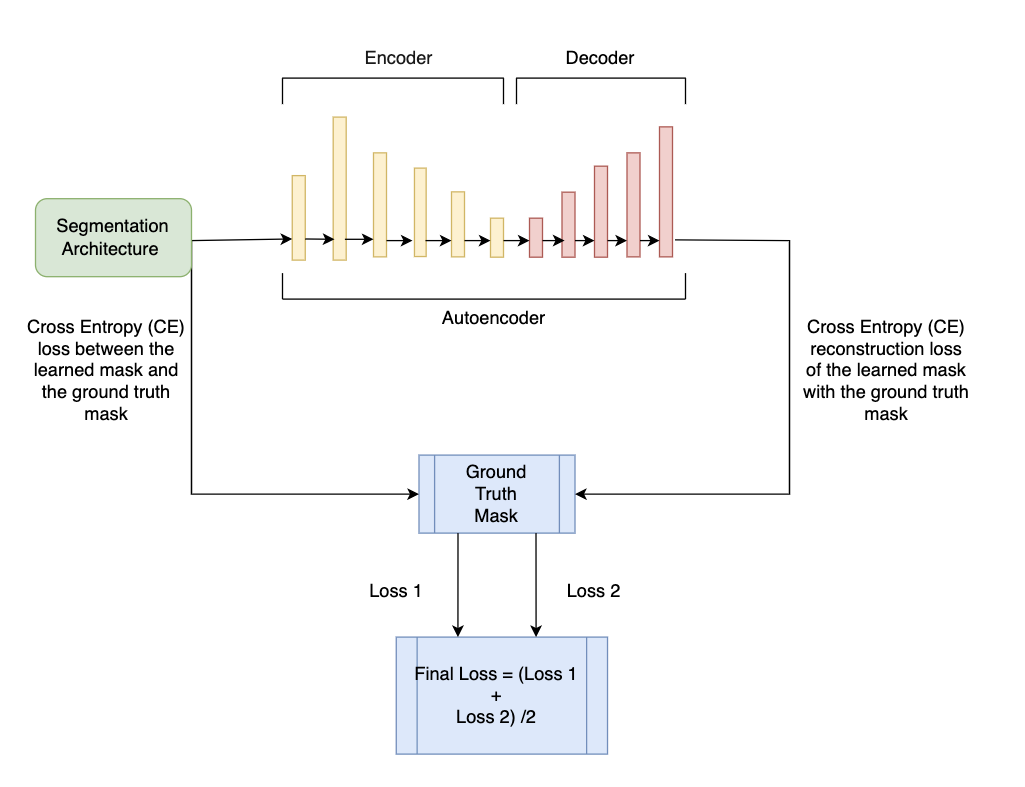}
\caption{Autoencoder Post-Processing (APP): we introduced the autoencoder labeled with yellow and red colour, after getting the segmentation mask from the segmentation architecture(in our case UNet and UNet++). }
\label{fig:ae}
\end{figure}
Figure~\ref{fig:ae} shows the post-processing autoencoder architecture in conjunction with segmentation architectures during training. The feedback from the autoencoder is only used during training and then the trained model is saved for inference. This incurs minimal computational costs during training (as shown in Section 5.1); the saved weights are no bigger in size than the saved weights without the autoencoder post-processing architecture and with no change in inference time.

\subsection{Classification}
We used their phenotype as markers to classify the different cell types. Wherein, the presence of a specific phenotype directly correlated to the presence of a cell type - T cell, B Cell, TFH-217, TFH-like cells and other cells. 
These cells’ presence or absence aided us in diagnosing dermatomyositis. As previously mentioned, to develop a better understanding of autoimmune diseases, the study of nonconforming cells to the mapped phenotype-cell classification is extremely important. We classified them as ’others’. This would be potentially helpful in diagnosis and understanding of novel cell patterns present in biopsies, which in-turn could help us understand autoimmune diseases to a better extent by categorizing novel phenotype-cell relationships. We used the auto-fluorescence images of size 352 by 469 and RGB. Since there could be multiple cells present per sample, this would be a multi-label classification. To address this class imbalance, we use the Focal loss \cite{lin2017focal} function that penalizes the dominant and the underrepresented class in a dataset. We used it to label distribution normalized class weights instead of vanilla cross-entropy, which fails to address this class imbalance. We also used a sixfold approach to train and reported our results on the test set to address any biases that might occur during training.
\begin{figure}[!ht]
\centering
\includegraphics[height=5cm]{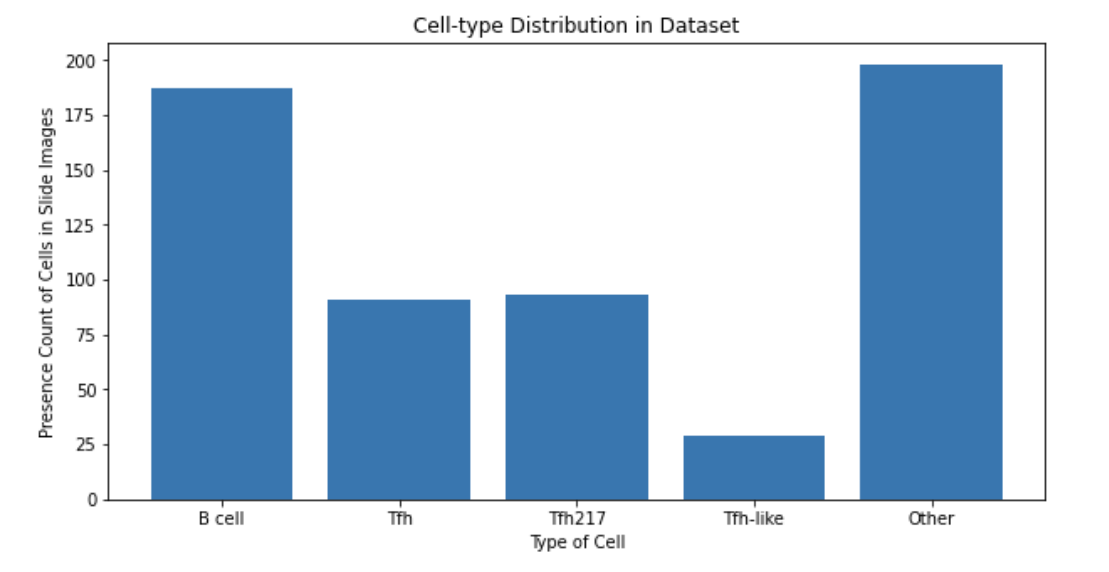}
\caption{Distribution of different cell types within the dataset. Observing that there is an imbalance in the distribution, we classified cells on the basis of the presence or absence of certain cell phenotype.}
\label{fig:example}
\end{figure}

\section{Experimental Details}
\subsection{Dataset}
We use the same dataset as used in \cite{VANBUREN2022113233}. This dataset contains 198 TIFF image samples, each containing slides of different protein-stained images - DAPI, CXCR3,CD19, CXCR5, PD1, CD4, CD27 and autofluorescence. Binary thresholds were set for each channel (1- DAPI, 2- CXCR3, 3- CD19, 4- CXCR5, 5- PD1, 6- CD4, 7- CD27, 8- Autofluorescence) to show presence/absence of each representative phenotype. These phenotypes were then classified into B cells and T cells using channels 2–7. The autofluorescence slides are an overlap of all the channels used for classifying cell types.

We use the DAPI stained image for semantic segmentation and the autofluorescence slide images for classification. This approach is a shift from previous work where researchers used DAPI channel images for both tasks as well as using UNet for segmentation and classification.

\subsection{Segmentation}
We use qubvel's implementation of Unet and Unet++ segmentation architectures \url{https://github.com/qubvel/segmentation_models.pytorch}. To start, we first convert the TIFF image file into NumPy and then into a PIL Image. We then apply Random Rotation, Random vertical and horizontal flip, and channel normalization before finally converting to tensors. We use the same splits for training, testing and validation as \cite{VANBUREN2022113233} to keep our results comparable. We use cross-entropy loss with component normalized weights, Adam optimizer with 1e-05 decay and cosine learning rate with minimum learning rate 3.4e-04. We also use \cite{hu2018squeeze} squeeze and excitation units in the decoder to add channel level attention.

For the Autoencoder processing architecture (as shown in Figure~\ref{fig:ae}), we use six layers, out of which five are downsampling layers followed by five upsampling layers. The encoder part contains 6 layers with first layer upscaling the input from 480 to 15360. From there on consecutive layers downsample from 15360 to 256, 128, 64, 32 to 16. This is then fed to the decoder which then systematically up scales it from 16, 32, 64, 128, 256 and 480. We use GELU (Gaussian Error Linear Unit) activation in all the layers. This post-processing architecture is only used during training as an added feedback mechanism; this helps the segmentation model improve learning. Times for saved model after training and inference remain same with or without the auto-encoder post-processing. We use Adam optimizer with a constant learning rate of 1e-3. For loss, we use the Cross Entropy (CE) loss function.

\subsection{Classification}
To address the class imbalance problem, we use focal loss \cite{lin2017focal}, which is essentially an oscillating cross entropy and this modulation fluctuates with easy and complex examples in the dataset.

We perform an 80/20 split for training to testing and use sixfold cross-validation. We then report the average F1-score across all the folds and different initialization in Section 5. We use Adam optimizer without weight decay with a cosine learning schedule and a learning rate of 1e-6 for 16 iterations. For a more generalized result and better optimization, we use Stochastic Weight Averaging \cite{izmailov2018averaging}. We use timm’s implementation \cite{rw2019timm} of CNN and Transformers for benchmarking.
\\

We conducted all our experiments on a single NVIDIA RTX-8000 GPU with 45GB of use-able video memory, 64GB RAM and two cores of CPU. We also used early-stopping with a patience of 5 epochs for both our classification and segmentation training pipeline. We compute average results over 5 runs with different seed values.

\section{Results and Discussion}

\subsection{Segmentation}

\subsubsection{Improvement over existing work.}
Since the images are 1408 by 1876, we tiled them in 256\textsuperscript{2} and then used blank padding at the edges to make them fit in size 256\textsuperscript{2}.

For segmentation, we started by using metrics, architecture and parameters as suggested in \cite{VANBUREN2022113233}. We begin with ImageNet weights as they are readily available for various backbones instead of the brain-MRI segmentation weights used in \cite{VANBUREN2022113233}. By changing the existing learning rate schedule from step to cosine and by using normalized weights of blank pixels and pixels that aren't blank (i.e., hold some information) in the cross-entropy loss, adding channel level attention with squeeze and excitation block to better feature map level channel relationships. We observed that the overall performance improved from 0.933 overall accuracy for \cite{VANBUREN2022113233} to 0.9834 for Unetplusplus with ResNet34 backbone – an improvement of 5\%.

\begin{figure}[!ht]
\centering
\includegraphics[height=5cm]{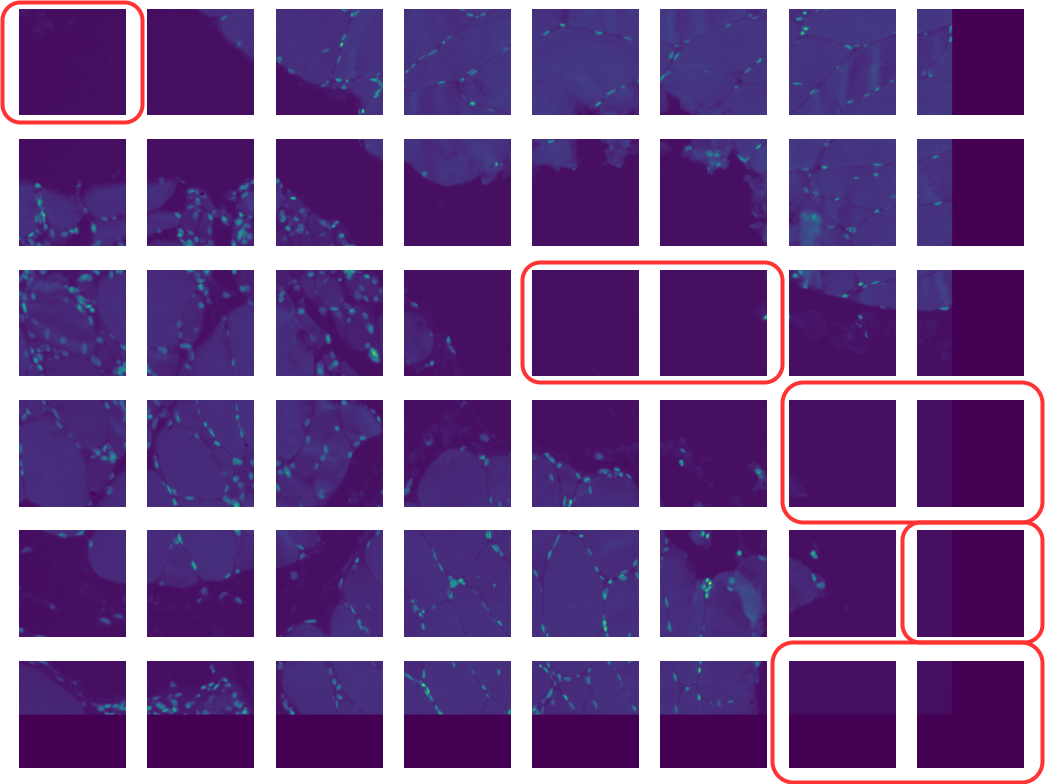}
\includegraphics[height=5cm]{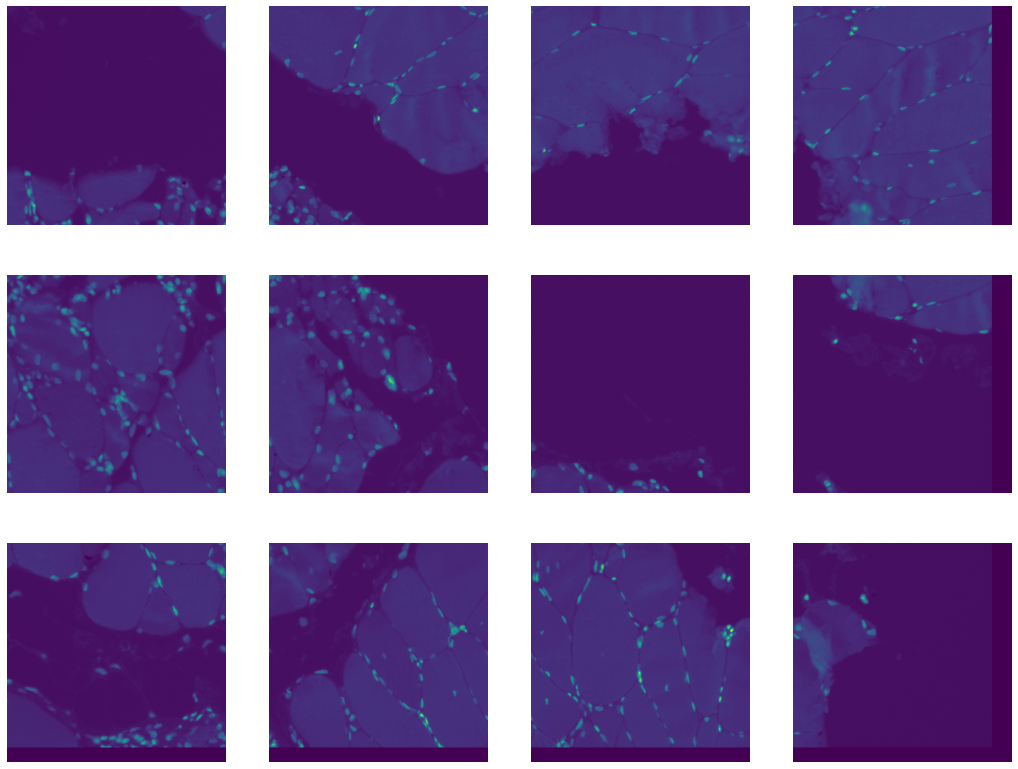}
\caption{Changing the image tiling reduces the number of blank tiles which are mostly concentrated towards the edges and some of the intermediate tiles. Top images shows 256\textsuperscript{2} tiling while bottom one shows 480\textsuperscript{2} tiling. As we can see for the top image tilled in 256 by 256 image size, the right bottom tiles are almost all empty. Including some of the tiles in the center portion. While for the bottom image tilled in 480 by 480 image size, no tiles are empty. The tiles highlighted in red boxes are empty.}
\label{fig:tile}
\end{figure}

We observed that: (i) by using image tiling of size 256\textsuperscript{2}, margin padding is primarily empty, and some of the tiles do not have any part of the cell stained on them. This might give false perception of performance as the model achieves higher metric performance from not learning anything. To mitigate this, we tiled the images into 480\textsuperscript{2}. This is depicted in Fig.\ref{fig:tile}; (ii) because accuracy is not the most appropriate metric for segmentation performance, we instead used IoU or Jaccrad Index as our metric. IoU is a better representation if the model has learned meaningful features.

\subsubsection{Fine-tuning.}

For further refining, we use one of the most cited and used architectures for biomedical segmentation, UNet \cite{ronneberger2015u}, and a nested version of UNet called the UNetplusplus \cite{zhou2018unet++}. These two architectures contain an encoder-decoder structure wherein the encoder is a feature extractor, and the decoder learns the segmentation mask on the extracted features. For the choice of the encoder, we used the ResNet family \cite{he2015deep} and the newer, more efficient family \cite{tan2019efficientnet}.
These architectures do not have an attention mechanism built-in. So we artificially included channel level attention mechanism using Squeeze and Excitation blocks \cite{hu2018squeeze}. In addition to these changes we propose adding a post-processing autoencoder architecture as described in Figure~\ref{fig:ae}.
\subsubsection{Adding Autoencoder.} As mentioned previously, we also experimented by adding autoencoder after the tuned segmentation architecture. We call this Autoencoder Post Processing or \textbf{APP}. 
We start by using the GELU (Gaussian Error Linear Unit) as our activation function for the autoencoder architecture (AE) as described in Section 3.1 and Figure~\ref{fig:ae}. We report the IoU score on test set for Unet and Unet++ in Table \ref{table:aechnage}. The autoencoder adds additional feedback to the segmentation architecture by computing reconstruction loss between the segmentation architecture’s output and ground truth. This adds some extra time during training but the space of saved model and inference time remain constant when compared to networks without the autoencoder post-processing architecture.

\subsubsection{Ablation Study.}
In the original implementation, we used GELU activation function with a constant learning rate and Adam optimizer in the autoencoder. In this section, we experimented with the hyperparameters of only the autoencoders to determine its effect on performance.
\begin{enumerate}
 
\item Computational cost: For a training set with 1452 images and validation set with 720 images of size 480\textsuperscript{2} over 50 epochs, we report the following testing time with and without using the GELU autoencoder post processing for UNet with ResNet backbones. We observed that there is an average increase of 1.875\% over the ResNet family of encoders.

\begin{table}[!ht]
\begin{center}
\caption{
This table shows training time for UNet with and without APP for ResNet family of encoders for 50 epochs. We observed that for smaller encoders like ResNet-18 and 34 the training time increase is greater as opposed to larger encoders.}
\label{table:computation}
\begin{tabular}{lll}
\hline\noalign{\smallskip}
\textbf{Encoder} & \textbf{Without APP} & \textbf{With APP} \\
\hline\noalign{\smallskip}
Resnet-18        & 1h 24m 44s           & 1h 27m 40s        \\
Resnet-34        & 1h 27m 41s           & 1h 30m 39s        \\
Resnet-50        & 1h 45m 45s           & 1h 46m 24s        \\
Resnet-101       & 1h 45m 24s           & 1h 46m 25s \\
\hline\noalign{\smallskip}
\end{tabular}
\end{center}
\end{table}
\item Using ReLU (Rectified Linear Unit) instead of GELU (Gaussian Error Linear Units) for activation function. Table~\ref{table:aechnage} shows the difference in performance for when we changed the activation of the autoencoders layers from ReLU to GELU. And for the same we observed that GELU performed better than ReLU activated autoencoder by around 3\% for UNet and UNet++. We observed in most cases for UNet and UNet++, the addition of the autoencoder post processing is highly beneficial. With the addition of autoencoder, GELU activation performed much better than ReLU activation. We further expand upon these results in Appendix 7.1.

\begin{table}[!ht]
\begin{center}
\caption{This table shows the IoU score on test set for UNet and UNet++ architectures with and without using cosine learning rate for the autoencoder. Except for ResNet-18 and 101 with UNet++, autoencoder always provide an improvement without cosine learning rate.}
\label{table:aechnage}
\begin{tabular}{lllllll}
\hline\noalign{\smallskip}
\textbf{Encoder} & \multicolumn{3}{c}{\textbf{UNet}}                                                                                                                                                                            & \multicolumn{3}{c}{\textbf{UNet++}}                                                                                                                                                                            \\
\textbf{}        & \textbf{\begin{tabular}[c]{@{}l@{}}Without \\ AE\end{tabular}} & \textbf{\begin{tabular}[c]{@{}l@{}}With \\ ReLU \\ AE\end{tabular}} & \textbf{\begin{tabular}[c]{@{}l@{}}With \\ GELU   \\ AE\end{tabular}} & \textbf{\begin{tabular}[c]{@{}l@{}}Without \\ AE\end{tabular}} & \textbf{\begin{tabular}[c]{@{}l@{}}With \\ ReLU    \\ AE\end{tabular}} & \textbf{\begin{tabular}[c]{@{}l@{}}With \\ GELU  \\ AE\end{tabular}} \\

\hline\noalign{\smallskip}
Resnet 18        & 0.4347                                                          & 0.4608                                                           & \textbf{0.4788}                                                  & \textbf{0.5274}                                                & 0.4177                                                           & 0.4707                                                           \\
Resnet 34        & 0.4774                                                         & 0.4467                                                           & \textbf{0.4983}                                                  & 0.3745                                                         & 0.4535                                                           & \textbf{0.4678}                                                  \\
Resnet 50        & 0.3798                                                         & \textbf{0.4187}                                                  & 0.3827                                                           & 0.4236                                                         & \textbf{0.4685}                                                  & 0.4422                                                           \\
Resnet 101       & 0.3718                                                         & 0.4074                                                           & \textbf{0.4402}                                                  & 0.4311                                                         & 0.4265                                                           & \textbf{0.4467}     \\
\hline\noalign{\smallskip}
\end{tabular}
\end{center}  
\end{table}

\item Using Adam optimizer with cosine learning schedule. For this we kept rest of the setup same. We only added a cosine learning schedule to the autoencoder architecture and ReLU activated layers. From Table~\ref{table:lrcosine} we observed that except for two cases in UNet++, autoencoder with constant learning rate perform much better than the one with Adman optimizer and cosine learning rate.

\begin{table}[!ht]
\begin{center}
\caption{
This table shows the IoU score on test set for UNet and UNet++ architectures. We compared results without, with autoencoder for both ReLU and GELU activations. Except for ResNet-18 with UNet++, autoencoder always provide an improvement. Within autoencoders we see that using GELU activation is much better than ReLU activation.}
\label{table:lrcosine}
\begin{tabular}{lllllll}
\hline\noalign{\smallskip}
\textbf{Encoder} & \multicolumn{2}{c}{\textbf{UNet}}                                                                                                                                                      & \multicolumn{2}{c}{\textbf{UNet++}}                                                                                                                                                    \\
\textbf{}        & \textbf{\begin{tabular}[c]{@{}l@{}}Without \\  Cosine \\ LR\end{tabular}} & \textbf{\begin{tabular}[c]{@{}l@{}}With Cosine \\ LR\end{tabular}} & \textbf{\begin{tabular}[c]{@{}l@{}}Without \\  Cosine \\ LR\end{tabular}} & \textbf{\begin{tabular}[c]{@{}l@{}}With  Cosine \\ LR\end{tabular}} \\
\hline\noalign{\smallskip}
Resnet 18        & \textbf{0.4608}                                                                             & 0.4106                                                                                   & 0.4177                                                                                      & \textbf{0.4717}                                                                          \\
Resnet 34        & \textbf{0.4467}                                                                             & 0.3665                                                                                   & \textbf{0.4535}                                                                             & 0.4345                                                                                   \\
Resnet 50        & \textbf{0.4187}                                                                             & 0.3965                                                                                   & \textbf{0.4685}                                                                             & 0.4268                                                                                   \\
Resnet 101       & \textbf{0.4074}                                                                             & 0.3846                                                                                   & 0.4265                                                                                      & \textbf{0.4518}    \\
\hline\noalign{\smallskip}
\end{tabular}
\end{center}
\end{table}

\end{enumerate} 

\subsection{Classification}
For classification our objective is to classify the different phenotypes in a given image and based on the presence of different cells. Since multiple labels could be assigned to the same image, this is a multi-class classification problem. From Fig.~\ref{fig:example}, we observed that the classes are highly imbalanced. \cite{VANBUREN2022113233} also reported an F1 score of 0.63 for the classification on this dataset. We improved upon their score with an F1 score of 0.891. This improvement could be attributed to the following, (i) we use focal loss \cite{lin2017focal} instead of cross entropy as the previous work. Focal loss applies a modulating term to the cross-entropy loss in order to focus learning on hard misclassified examples. It is a dynamically scaled cross-entropy loss, where the scaling factor decays to zero as confidence in the correct class increases. Intuitively, this scaling factor can automatically down-weight the contribution of easy examples during training and rapidly focus the model on hard examples. (ii) We used newer and more efficient architectures. We benchmarked pure CNNs, Transformers and newer generation of CNNs trained with Transformer-like techniques Convnext\cite{liu2022convnet}. As mentioned in Section 3, in this section we present our results and discuss the effect of different architectures.

\subsubsection{Effect of Architecture.} 
Recently, the emergence of Transformers for vision tasks has taken the field of computer vision by storm. They have been able to improve upon many benchmarks set by CNNs. They are particularly of interest in the medical imaging because of their global fidelity as opposed to the centered fidelity of CNNs.

More recently, some of the Transformer techniques have also been applied to train CNNs. This has given rise to new architectures like the ConvNeXts family of pure ConvNet models \cite{liu2022convnet}. We studied the effect of using different architectures on performance and hence have found the most suitable architecture for the multiclass classification task at hand. We resized each image to be 384\textsuperscript{2}.
\begin{table}[]
\begin{center}
\caption{We report the F1 score for with ImageNet initialization for latest set of CNNs and Transformers. We observed the usual trend of increasing ImageNet performance with increasing size of the model is not followed. Overall the best performance is achieved by nfnet-f3 for CNNs. Overall, out of all the tested architectures Swin Transformer Base with Patch 4 and Window 12 performed the best}
\label{table:allcnn}
\begin{tabular}{ll}
\hline\noalign{\smallskip}
\textbf{Model}    & \textbf{Test F1-Score} \\
\hline\noalign{\smallskip}
Resnet-18         & 0.8635±0.0087        \\
Resnet-34         & 0.82765±0.0073      \\
Resnet-50         & 0.8499±0.007           \\
Resnet-101        & 0.871±0.009            \\
Efficient-net B0  & 0.8372±0.0007         \\
Efficient-net B1  & 0.8346±0.0026         \\
Efficient-net B2  & 0.828±0.00074          \\
Efficient-net B3  & 0.8369±0.0094          \\
Efficient-net B4  & 0.8418±0.0009          \\
Efficient-net B5  & 0.8463±0.00036         \\
Efficient-net B6  & 0.8263±0.00147         \\
Efficient-net B7  & 0.8129±0.001           \\
nfnet-f0          & 0.82035±0.007          \\
nfnet-f1          & 0.834±0.007            \\
nfnet-f2          & 0.8652±0.0089          \\
nfnet-f3          & 0.8898±0.0011          \\
nfnet-f4          & 0.8848±0.0109 \\
nfnet-f5          & 0.8161±0.0074          \\
nfnet-f6          & 0.8378±0.007          \\
ConvNext-tiny     & 0.81355±0.0032          \\
ConvNext-small    & 0.84795±0.00246        \\
ConvNext-base     & 0.80675±0.002          \\
ConvNext-large    & 0.8452±0.000545        \\
Swin Transformer large\\ (Patch 4 Window 12)\ & 0.8839±0.001           \\
Swin Transformer Base\\ (Patch 4 Window 12)\  & \textbf{0.891±0.0007}           \\
Vit-Base/16                                                                          & 0.8426±0.007           \\
Vit-Base/32                                                                          & 0.8507±0.0079        \\
Vit-large/16                                                                         & 0.80495±0.0077        \\
Vit-large/32                                                                         & 0.845±0.0077          \\
\hline\noalign{\smallskip}
\end{tabular}
\end{center}
\end{table}

\begin{table}[]
\begin{center}
\caption{We compare the best results provided by our algorithm (in bold) as compared to previous benchmark on the same dataset.}
\label{table:vbcmp}
\begin{tabular}{ll}
\hline\noalign{\smallskip}
\textbf{Model}                                                                       & \textbf{Test F1-Score} \\
\hline\noalign{\smallskip}
\begin{tabular}[c]{@{}l@{}}Swin Transformer Large\\ (Patch 4 Window 12)\end{tabular} & \textbf{0.891±0.0007}            \\
nfnet-f3          & \textbf{0.8898±0.0109}       \\
Vanburen et all\cite{VANBUREN2022113233} &0.63 \\
\hline\noalign{\smallskip}
\end{tabular}
\end{center}
\end{table}


Amongst CNNs and Transformers, the peak performance is closely matched with Swin Transformer Base (Patch 4 Window 12) achieving a new state of the art performance for the autoimmune dataset with an F1 Score of 0.891, an improvement of 26.1\% over previous work by \cite{VANBUREN2022113233}; compared to nfnet-f3 that provides peak performance for CNNs with an F1 score of 0.8898.

\section{Conclusion}
Our framework can be adapted to other tissues and diseases datasets. It provides an efficient approach for clinicians to identify and detect cells within histopathology images in order to better comprehend the architecture of inflammation (i.e., which cell types are involved in inflammation at the tissue level, and how cells interact with one other).
\\
Based on our experimentation, we observe that for segmentation of biopsies affected by dermatomyositis, it is better to use Imagenet initialization with a normalized cross-entropy loss. Further performance can be increased by using our proposed autoencoder post-processing architecture (APP). APP gains 3\% consistently over architectures without any post-processing segmentation architecture. This addition comes at minimal extra training cost and at no extra space and time to develop inference models.\\
For classification, using stochastic weight averaging improves generalization and class normalized weights with focal loss and helps to counter the class imbalance problem. In comparing architectures, the performance is relatively similar for CNNs and transformers, but transformers perform slightly better than CNNs. These changes helped us register an improved performance of 8\% in segmentation and of 26\% in classification performance on our dermatomyositis dataset.\\

{\bfseries\noindent Acknowledgment.} We would like to thank NYU HPC team for assisting us with our computational needs. We would also like to thank Prof. Elena Sizikova (Moore Sloan Faculty Fellow, Center for Data Science (CDS), New York University (NYU)) for her valuable feedback.


\clearpage
%
%
\bibliographystyle{splncs04}
\bibliography{egbib}
\clearpage
\section{Appendix}

\subsection{Expansion of Results}

In Tables~\ref{table:unetfull} and \ref{table:unetppfull} we show complete results with mean and standard deviation. These are an expansion of Table 2 in Section 5.1 of the main paper. Tables were compressed to save space and only focus on the main results. To provide a complete picture, we added extended results in this section. 

\begin{table}[!ht]
\begin{center}
\caption{This table shows the IoU score on the test set for UNet. We compared results without and with autoencoder for both ReLU and GELU activations for UNet Architecture. These results are averaged over five runs with different seed values. We observed that in all cases addition of APP improved performance. GELU activated APP seems out perform the ReLU activated APP in all cases except for ResNet-50.}
\label{table:unetfull}
\begin{tabular}{llll}
\hline\noalign{\smallskip}
\textbf{Encoder} & \multicolumn{3}{c}{\textbf{UNet}}                                                                                                                                                                                                                                                                                \\
\textbf{}        & \textbf{\begin{tabular}[c]{@{}l@{}}Without \\ AE\end{tabular}} & \textbf{\begin{tabular}[c]{@{}l@{}}With \\ ReLU \\ AE\end{tabular}} & \textbf{\begin{tabular}[c]{@{}l@{}}With \\ GELU   \\ AE\end{tabular}} \\

\hline\noalign{\smallskip}
ResNet 18        & 0.4347±0.0006                                                          & 0.4608±0.0001                                                           & \textbf{0.4788±0.0004}                                                                                                           \\
ResNet 34        & 0.4774±0.0004                                                          & 0.4467±0.0012                                                            & \textbf{0.4983±0.0008}                                                                                                 \\
ResNet 50        & 0.3798±0.00072                                                          & \textbf{0.4187±0.0006 }                                                  & 0.3827±0.0003                                                           \\
ResNet 101       & 0.3718±0.0001                                                          & 0.4074±0.0012                                                           & \textbf{0.4402±0.00018 }                                                      \\
\hline\noalign{\smallskip}
\end{tabular}
\end{center}  
\end{table}

\begin{table}[!ht]
\begin{center}
\caption{This table shows the IoU score on the test set for UNet++. These results are averaged over five runs with different seed values. We compare results without and with autoencoder for both ReLU and GELU activations for UNet++ Architecture. We observed that in most cases, APP improves performance except for UNet++ with Resnet-18, where APP segmentation techniques lag by around 5\%. However, as a counter for ResNet-34 APP-based segmentation techniques are almost 10\% better than UNet++ without APP.}
\label{table:unetppfull}
\begin{tabular}{llll}
\hline\noalign{\smallskip}
\textbf{Encoder}                                                                                                                                                                           & \multicolumn{3}{c}{\textbf{UNet++}}                                                                                                                                                                            \\
\textbf{}        &  \textbf{\begin{tabular}[c]{@{}l@{}}Without \\ AE\end{tabular}} & \textbf{\begin{tabular}[c]{@{}l@{}}With \\ ReLU    \\ AE\end{tabular}} & \textbf{\begin{tabular}[c]{@{}l@{}}With \\ GELU  \\ AE\end{tabular}} \\

\hline\noalign{\smallskip}
ResNet 18                                                        & \textbf{0.5274±0.0004}                                                & 0.4177±0.0005                                                           & 0.4707±0.00067                                                           \\
ResNet 34                                                        & 0.3745±0.0006                                                         & 0.4535±0.0008                                                           & \textbf{0.4678±0.0004}                                                  \\
ResNet 50                                                               & 0.4236±0.0004                                                         & \textbf{0.4685±0.0002}                                                  & 0.4422±0.0007                                                           \\
ResNet 101                                                       & 0.4311±0.0003                                                         & 0.4265±0.0002                                                          & \textbf{0.4467±0.0003}     \\
\hline\noalign{\smallskip}
\end{tabular}
\end{center}  
\end{table}

\subsection{Autoencoder with efficientnet encoder for segmentation}

In Table~\ref{table:3} and \ref{table:4} we compared the time taken to train and the performance of the respective trained architecture for segmentation using EfficientNet encoders. We observed that with the addition of autoencoder post-processing, training time increased by an average of 3m 7.3s over 50 epochs (averaged over the entire efficientnet family). This is an increase of 2.93\% in training time over the eight encoders. In other words, an average increase of 0.36\% increase in time per encoder over 50 epochs.

Performance wise architecture with autoencoder post-processing consistently outperformed segmentation architectures without them by 2.75\%.

\begin{table}[]
 \begin{center}
 \caption{In this table we report the running time averaged over 5 runs with different seeds, for efficient-net encoder family with UNet.The variation is almost negligible($<6s$).}
 \label{table:3}
\begin{tabular}{lll}
\hline\noalign{\smallskip}
\textbf{Encoder} & \textbf{Without APP} & \textbf{With APP} \\
\hline\noalign{\smallskip}
B0               & 1h 26m 27s           & 1h 29m 04s        \\
B1               & 1h 31m 16s           & 1h 33m 42s        \\
B2               & 1h 32m 12s           & 1h 34m 27s        \\
B3               & 1h 38m            & 1h 40m 33s
\\
B4               & 1h 44m 20s            & 1h 50m 02s
\\
B5               & 1h 55m 46s           & 1h 58m 40s
\\
B6               & 2h 06m 55s           & 2h 10m 08s
\\
B7               & 2h 16m 40s            & 2h 19m 59s
\\
\hline\noalign{\smallskip}
\end{tabular}
\end{center}
\end{table}

\begin{table}[]

 \begin{center}
 \caption{In this table we report the IoU averaged over 5 runs with different seeds, for efficient-net encoder family with UNet architecture.}
 \label{table:4}
\begin{tabular}{lll}
\hline\noalign{\smallskip}
\textbf{Encoder} & \textbf{Without APP} & \textbf{With APP} \\
\hline\noalign{\smallskip}
B0               &0.3785±0.00061            &\textbf{0.4282±0.0008}       \\
B1               &0.3301±0.0002          &\textbf{0.4237±0.0006}       \\
B2               &0.2235±0.0007         &\textbf{0.3735±0.0009}      \\
B3               &\textbf{0.3982±0.0007}           &0.2411±0.0004
\\
B4               &0.3826±0.0004          &\textbf{0.3829±0.0006}
\\
B5               &0.4056±0.0008         &\textbf{0.4336±0.0008 }
\\
B6               &0.4001±0.0001        &\textbf{0.4311±0.0006}
\\
B7               &0.3631±0.0002          &\textbf{0.3937±0.0004}
\\
\hline\noalign{\smallskip}
\end{tabular}
\end{center}
\end{table}

Similarly, we compared computational and performance for UNet++ with and without the autoencoder post-processing in Tables~\ref{table:5}
and \ref{table:6} respectively. In this case, we observed that the gain in performance with autoencoder post-processing is 5\% averaged over the efficientnet family of encoders. This also corresponds to a 3m 7s increase in training time which is an increase of 2.6\%.
\begin{table}[]
\label{table:unetppcompeffnet}
 \begin{center}
 \caption{In this table we report the running time averaged over 5 runs with different seeds, for efficient-net encoder family with UNet++.}
  \label{table:5}
\begin{tabular}{lll}
\hline\noalign{\smallskip}
\textbf{Encoder} & \textbf{Without APP} & \textbf{With APP} \\
\hline\noalign{\smallskip}
B0               & 1h 32m 50s           & 1h 35m 31s        \\
B1               & 1h 37m 40s           & 1h 40m 51s        \\
B2               & 1h 38m 30s           & 1h 40m 41s        \\
B3               & 1h 46m 30s            & 1h 49m 34s
\\
B4               & 1h 54m 01s            & 1h 57m 41s
\\
B5               & 2h 07m 54s           & 2h 11m 39s
\\
B6               & 2h 20m 23s           & 2h 23m 41s
\\
B7               & 2h 29m 01s            & 2h 32m 04s
\\
\hline\noalign{\smallskip}
\end{tabular}
\end{center}
\end{table}

\begin{table}[]
\label{table:unetpppereffnet}
 \begin{center}
 \caption{In this table we report the IoU averaged over 5 runs with different seeds, for efficient-net encoder family with UNet++ architecture.}
 \label{table:6}
\begin{tabular}{lll}
\hline\noalign{\smallskip}
\textbf{Encoder} & \textbf{Without APP} & \textbf{With APP} \\
\hline\noalign{\smallskip}
B0               &0.3584±0.0002            &\textbf{0.3751±0.0007}      \\
B1               &0.4260±0.0005           &\textbf{0.4269±0.0003}       \\
B2               &0.3778±0.0007          &\textbf{0.3942±0.0009}        \\
B3               &0.3928±0.0006            &\textbf{0.4174±0.0003}
\\
B4               &0.4138±0.0003           &\textbf{0.4273±0.0002}
\\
B5               &\textbf{0.3884±0.0001}           &0.3875±0.0005
\\
B6               &0.4090±0.0008        &\textbf{0.4214±0.0007}
\\
B7               &0.3784±0.0009         &\textbf{0.4002±0.0005} 
\\
\hline\noalign{\smallskip}
\end{tabular}
\end{center}
\end{table}

\subsection{Metrics Description}

For measuring segmentation performance, we use IoU or intersection over union metric. It helps us understand how similar sample sets are.\\ 
\begin{center}
$IoU = \frac{\text{area of overlap}}{\text{area of union}}=
\frac{
    \tikz{\fill[draw=blue, very thick, fill=red!5] (0,0) rectangle (2,2) (0.5,-0.5) rectangle (2.5,1.5);
    \fill[draw=red, very thick, fill=white, even odd rule] (0,0) rectangle (2,2) (0.5,-0.5) rectangle (2.5,1.5);}}
{\tikz{\fill[draw=red, fill=red!5, very thick] (0,0) rectangle (2,2) (0.5,-0.5) rectangle (2.5,1.5);}}$
\end{center}

Here the comparison is made between the output mask by segmentation pipeline against the ground truth mask.

For measuring classification performance, we use the F1 score.\\
Computed as F1 = $\frac{\text{2*Precision*Recall}}{\text{Precision+Recall}} = \frac{\text{2*TP}}{\text{2*TP+FP+FN}}$

\subsection{Effect of different weights} ImageNet initialization has been the defacto norm for most transfer learning tasks. Although in some cases, as in \cite{agarwal2022classification} it was observed that noisy student weights performed better than ImageNet initialization. To study the effect in our case, we used advprop and noisy student initialization. ImageNet weights for initialization work for medical data not because of feature reuse but because of better weight scaling and faster convergence \cite{raghu2019transfusion}. Noisy student training \cite{xie2020self} extends the idea of self-training and distillation with the use of equal-or-larger student models, and noise such as dropout, stochastic depth,
and data augmentation via RandAugment is added to the student during learning so
that the student generalizes better than the teacher. First, an EfficientNet model is trained on labelled images and is used as a teacher to generate pseudo labels for 300M unlabeled images. We then train a larger EfficientNet as a student model on the combination of labelled and
pseudo-labelled images. This helps reduce the error rate, increases robustness and improves performance over the existing state-of-the-art on ImageNet. 

(ii)AdvProp training, which banks on Adversarial examples, which are commonly viewed as a threat
to ConvNets. In \cite{xie2020adversarial} they present an opposite perspective: adversarial examples can be used to improve image recognition models. They treat adversarial examples as additional examples to prevent overfitting. It performs better
when the models are bigger. This improves upon performance for various ImageNet and its' subset benchmarks. 

Since initially all these were developed for the  EfficientNet family of the encoders, we used them for benchmarking. We present their results in Table~\ref{table:segweights}.

\begin{table}[!h]
\begin{center}
\caption{Using different initialization, we saw that the performance of different encoders of the EfficientNet family on UNet. We report the IoU over the test set in the following table.
We observe that while performance gains for smaller models, ImageNet initialisation works better for larger models. Also, the fact that advprop and noisy are not readily available for all models, hence the choice of ImageNet still dominates.}
\label{table:segweights}

\begin{tabular}{llll}
\hline\noalign{\smallskip}
\textbf{Encoder} & \textbf{ImageNet} & \textbf{Advprop} & \textbf{Nosiy}  \\
\noalign{\smallskip}
\hline
\noalign{\smallskip}
B0               & 0.3785±0.00061             & 0.3895±0.001            & \textbf{0.4081±0.0006} \\
B1               & 0.3301±0.0002            & 0.2330±0.0006           & \textbf{0.3817±0.0008} \\
B2               & 0.2235±0.0007            & 0.3823±0.0004           & \textbf{0.4154±0.0001 } \\
B3               & \textbf{0.3982±0.0007}   & 0.3722±0.0004           & 0.3509±0.0007           \\
B4               & 0.3819±0.0004             & \textbf{0.4366±0.0006}  & 0.4333±0.0003           \\
B5               & \textbf{0.4056±0.0008}   & 0.3910±0.0001           & 0.3806±0.0004           \\
B6               & \textbf{0.4001±0.0001 }   &  0.3807±0.0005                 & 0.3941±0.0006           \\
B7               &  \textbf{0.3631±0.0002}                 & 0.3543±0.0009                 & 0.3308±0.0006                \\
\hline
\end{tabular}
\end{center}
\end{table}

Similarly, we conduct similar experiments for classification with different initialization. We reported these results in Table \ref{table:effnet}
 
 \begin{table}[]
 \begin{center}
\caption{We report the F1 score for different initializations for the EfficientNet family of encoders. We reported the average of 6-fold runs on the test set with five different seed values. We observed that 0.8463 is the peak with ImageNet, 0.843 with advprop and 0.8457 with noisy student initialization.}
\label{table:effnet}
\begin{tabular}{llll}
\hline\noalign{\smallskip}
\textbf{Encoder} & \textbf{ImageNet}       & \textbf{Advprop}       & \textbf{Noisy Student} \\
\hline\noalign{\smallskip}
B0               & 0.8372± 0.0007          & \textbf{0.8416±0.0008} & 0.839±0.0008           \\
B1               & 0.8346±0.0026           & 0.8367±0.0007          & \textbf{0.8448±0.0014} \\
B2               & 0.828±0.00074           & \textbf{0.843±0.00138} & 0.8430±0.0012          \\
B3               & 0.8369±0.0094           & 0.84138±0.00135        & \textbf{0.8457±0.007}  \\
B4               & \textbf{0.8418±0.0009}  & 0.82135±0.00058        & 0.8377±0.00032         \\
B5               & \textbf{0.8463±0.00036} & 0.8133±0.002           & 0.8326±0.00066         \\
B6               & \textbf{0.8263±0.00147} & 0.8237±0.004           & 0.8233±0.0065          \\
B7               & 0.8129±0.001            & 0.8132±0.004           & \textbf{0.8257±0.0008} \\
\hline\noalign{\smallskip}
\end{tabular}
\end{center}
\end{table}

As we can see for segmentation, ImageNet initialization performed better in most cases. Similarly, in classification, it not only performed better in most cases but also provided the best overall result—these inferences, combined with the fact that advprop and noisy student requires additional computational resources. Hence we decide to stick with ImageNet initialization.

\subsection{Expansion on Experimental Details}
\subsubsection{Segmentation}

We used PyTorch lightning's \cite{falcon2019pytorch} seed everything functionality to seed all the generator values uniformly. For setting the seed value, we randomly generated a set of numbers in the range of 1 and 1000. We did not perform an extensive search of space to optimise performance with seed value as suggested in \cite{DBLP:journals/corr/abs-2109-08203}. We used seed values 26, 77, 334, 517 and 994. For augmentation, we used conversion to PIL Image to apply random rotation (degrees=3), random vertical and horizontal flip, then conversion to tensor and finally channel normalisation.
We could have used a resize function to reshape the 1408 by 1876 Whole Slide Images (WSI), but we instead tilled them in 480 square tile images.
We then split them into a batch size of 16 before finally passing through the segmentation architecture (UNet/UNet++). We used channel attention only decoder, with ImageNet initialisation and a decoder depth of 3 (256, 128, 64).

We used cross-entropy loss with dark/light pixel normalization, Adam optimizer with LR set to 3.6e-04 and weight decay of 1e-05. We used a cosine scheduling rate with a minimum value set to 3.4e-04. 

\subsubsection{APP Segmentation} When using APP we used GELU activation by default with adam optimizer and lr set to 1e-3.

\subsubsection{Classification} For Classification, we used the same seed values with PyTorch lightning's \cite{falcon2019pytorch} seed everything functionality, as described for segmentation above. For augmentation, we resized the images to 384 square images, followed by randomly applying colour jitter (0.2, 0.2, 0.2) or random perspective (distortion scale=0.2) with probability 0.3, colour jittering (0.2, 0.2, 0.2) or random affine (degrees=10) with probability 0.3, random vertical flip and random horizontal flip with probability 0.3 and finally channel normalization. 

We used Stochastic weigh averaging with adam optimizer. We used a cosine learning rate starting at 1e-3 and a minimum set to 1e-6. We used focal loss with normalized class weight as our loss function. We used 6-fold validation with each fold of 20 epochs and batch size of 16. We used same parameters for both CNN and Transformers. 
\end{document}